\documentclass[12pt]{revtex4}
  \usepackage{graphicx}

\begin{document}

\title{Meissner response of  a bulk superconductor with an embedded sheet of reduced penetration depth}

\author{J. R. Kirtley$^{1,2,3,*}$, B. Kalisky$^{1,2}$, L. Luan$^{1,4}$, and K.A. Moler$^{1,2,4}$}

\address{1. Geballe Laboratory for Advanced Materials, Stanford University, Stanford, California 94305-4045, USA}
\address{2. Department of Applied Physics, Stanford University, Stanford, California 94305-4045, USA}
\address{3. IBM Watson Research Center, Route 134 Yorktown Heights, NY 10598, USA}
\address{4. Department of Physics, Stanford University, Stanford, California, 94305-4045, USA}

%\ead{jkirtley@stanford.edu}
\begin{abstract}
We calculate the change in susceptibility resulting from a thin sheet with reduced penetration depth embedded  perpendicular to the surface of an isotropic superconductor, in a geometry applicable to scanning Superconducting QUantum Interference Device (SQUID) microscopy, by numerically solving Maxwell's and London's equations using the finite element method. The predicted stripes in susceptibility agree well in shape with the observations of Kalisky {\it et al.}\cite{kalisky2009} of enhanced susceptibility above twin planes in the underdoped pnictide superconductor Ba(Fe$_{1-x}$Co$_x$)$_2$As$_2$ (Ba-122). By comparing the predicted stripe amplitudes with experiment and using the London relation between penetration depth and superfluid density, we estimate the enhanced  Cooper pair  density on the twin planes, and the barrier force for a vortex to cross a twin plane.  Fits to the observed temperature dependence of the stripe amplitude suggest that  the twin planes have a higher critical temperature  than the bulk, although stripes are not observed above the bulk critical temperature. 
\end{abstract}

\maketitle

\section{Introduction}

Scanning SQUID microscopy has been used very successfully to elucidate fundamental properties of superconductors from spatially resolved magnetometry images, including the determination of the pairing symmetry of the cuprate superconductors \cite{tsuei1994}, tests of the interlayer tunneling model for high temperature superconductivity \cite{moler1998}, and the placement of limits on spin-charge separation in high temperature superconductors \cite{bonn2001}. Recently a new dimension has been added to scanning SQUID measurements: scanning SQUID susceptometry  \cite{gardner2001} has enabled spatially resolved measurements of superconducting penetration depths \cite{tafuri2004, hicks2009}, the observation of spontaneous persistent currents in single mesoscopic normal rings \cite{bluhm2009}, and measurements of fluctuations in single mesoscopic superconducting rings \cite{koshnick2007}. An advantage of scanning SQUID susceptibility measurements, aside from their exceptional sensitivity, is that they can be easily, reliably, and quantitatively modeled \cite{kogan2003}. Recently Kalisky {\it et al.} \cite{kalisky2009} reported stripes of enhanced susceptibility in scanning SQUID susceptometer measurements of single crystals of the pnictide superconductor Ba(Fe$_{1-x}$Co$_x$)$_x$As$_2$. They associated these stripes with twin boundaries in the superconductor, for a number of reasons: 1) The stripe spacings and orientation were consistent with optical and x-ray observations of stripes in the same \cite{kalisky2009} and similar \cite{tanatar2009} samples. 2) The stripes were only observed for underdoped samples, which undergo a tetragonal to orthorhombic crystal structure transition and therefore have twins, but not for optimally doped or overdoped samples. 3) The stripes changed position when the samples were warmed above the tetragonal-orthorhombic transition temperature, but not when the sample was only warmed above the superconducting transition temperature. Further, it was observed that vortices did not pin on the stripes, and that when dragged using either a SQUID sensor, or a magnetic force microscope tip, the vortices did not cross the stripes. Enhanced diamagnetic susceptibility and enhanced critical temperatures associated with twinning have been reported previously from bulk measurements of several elemental superconductors \cite{khlyustikov1987}. 

The observation of stripes in susceptibility is quite interesting qualitatively, because it indicates that the superconductivity is different on the twin boundaries than in the bulk in these novel superconductors. However, in order to fully understand these results it is important to model them quantitatively. This is a challenge,  because of the special sample geometry involved. Kogan \cite{kogan2003} developed a theory for the Meissner response of anisotropic superconductors to several types of locally applied magnetic fields, including from a circular field coil. This theory produces exact solutions for the problem of scanning SQUID susceptometry of a homogeneous superconductor. However, it is not immediately apparent how to apply this theory to our geometry. In the present paper we use numerical methods to solve the problem of local susceptibility measurements of an inhomogeneous superconductor with a planar sheet with different superconducting properties oriented perpendicular to the bulk surface. We apply the results of this modeling to the experiments of Kalisky {\it et al.} and find good agreement. Our analysis indicates that there is substantial additional Cooper pair density on the twin boundaries, and that the critical temperature of the twin boundary region is higher than that in the bulk.

\section{Enhanced susceptibility images} 

\begin{figure}
 \includegraphics[width=6in]{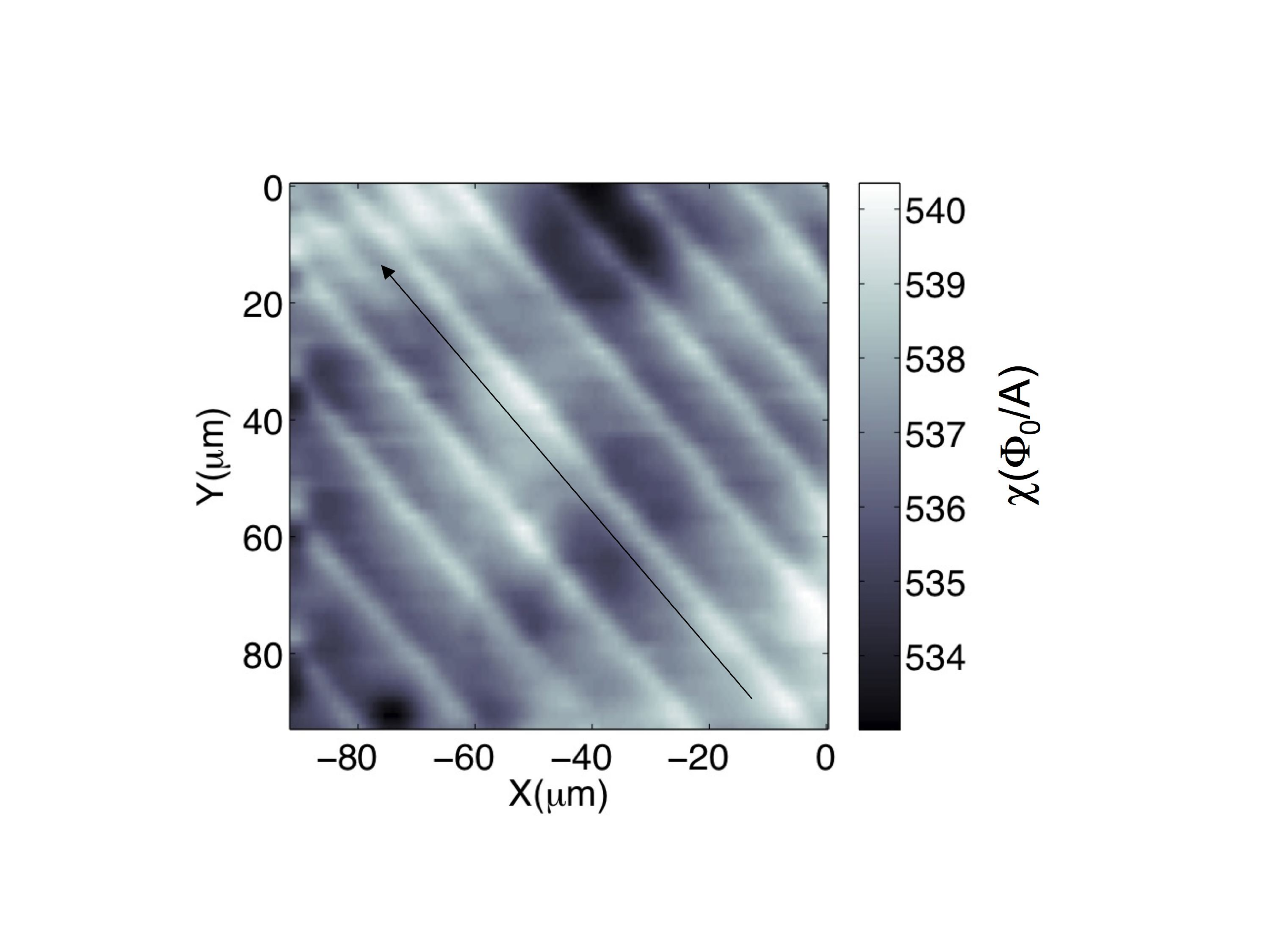}
\caption{Scanning SQUID susceptibility image of an underdoped ($x$=0.051, T$_c$=18.25K) single crystal of Ba(Fe$_{1-x}$Co$_x$)$_2$As$_2$ at T=17K. The arrow indicates the direction along which the stripes were averaged to form the cross-section plotted in Figure \ref{fig:cross_ave}.}.
\label{fig:cross_image}
\end{figure}

\begin{figure}
\includegraphics[width=4in]{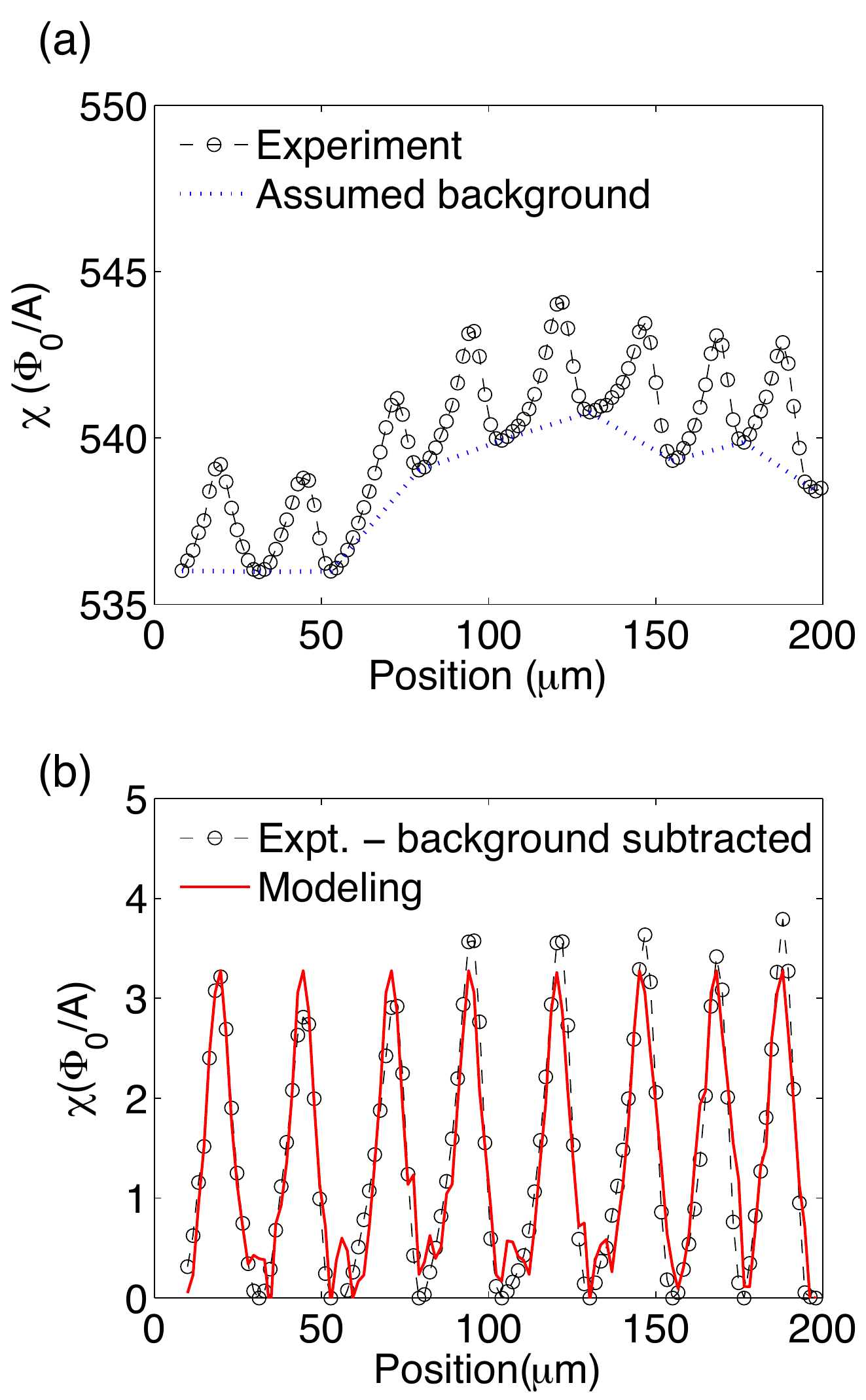}
\caption{(a) Averaged cross section along the stripes direction of the bottom 3/5 of the data of Figure \ref{fig:cross_image} (symbols, dashed line). Subtracting out the piece-wise linear background (dotted line in (a)) results in the symbols and dashed line in (b). The solid line in (b) is modeling as described in the text. }
\label{fig:cross_ave}
\end{figure}

 An example of the stripes of enhanced susceptibility observed by Kalisky {\it et al.} is shown in Figure \ref{fig:cross_image}. A scanning SQUID susceptometer sensor has a single turn field coil surrounding a co-planar pickup loop integrated into the SQUID sensor. Susceptometry measurements are made by applying a small alternating current to the field coil and phase sensitively sensing the response of the SQUID, proportional to the flux through the field coil, to this current. The data shown in Figure \ref{fig:cross_image} was taken at $T$=17 K using a sensor \cite{huber2008} with an effective field coil radius of $R$=8.85$\mu$m and an effective pickup loop radius of $r$=2.1$\mu$m. The effective height of the sensor above the sample surface was $z_0$=1.5$\mu$m, derived from fitting magnetometry images of individual superconducting vortices in the sample. The $ac$ current through the SQUID was 0.25 mA, corresponding to a  magnetic induction at the sample surface of 17$\mu$T. The sign convention is chosen such that diamagnetic shielding is positive: higher numbers and white colors represent stronger diamagnetic shielding.

In order to compare experimental data with our modeling, we averaged the image in Figure \ref{fig:cross_image} along the direction indicated by the arrow to obtain the cross-section displayed in Figure \ref{fig:cross_ave}a. There are broad spatial variations in the susceptibility in addition to the stripes. We subtract them from the data using a piece-wise linear background indicated by the dashed line in Fig. \ref{fig:cross_ave}a. This results in the data displayed as symbols in Fig. \ref{fig:cross_ave}b. The susceptibility peaks in this cross-sectional average have  amplitudes of $3.37\pm0.35 \Phi_0/A$ and full widths at half-maximum of $8.92\pm0.62\mu m$. This is to be compared with a low temperature susceptibility of about 600 $\Phi_0/A$. The solid line in Figure \ref{fig:cross_ave} is the result of modeling as described below.

\begin{figure}
\includegraphics[width=6in]{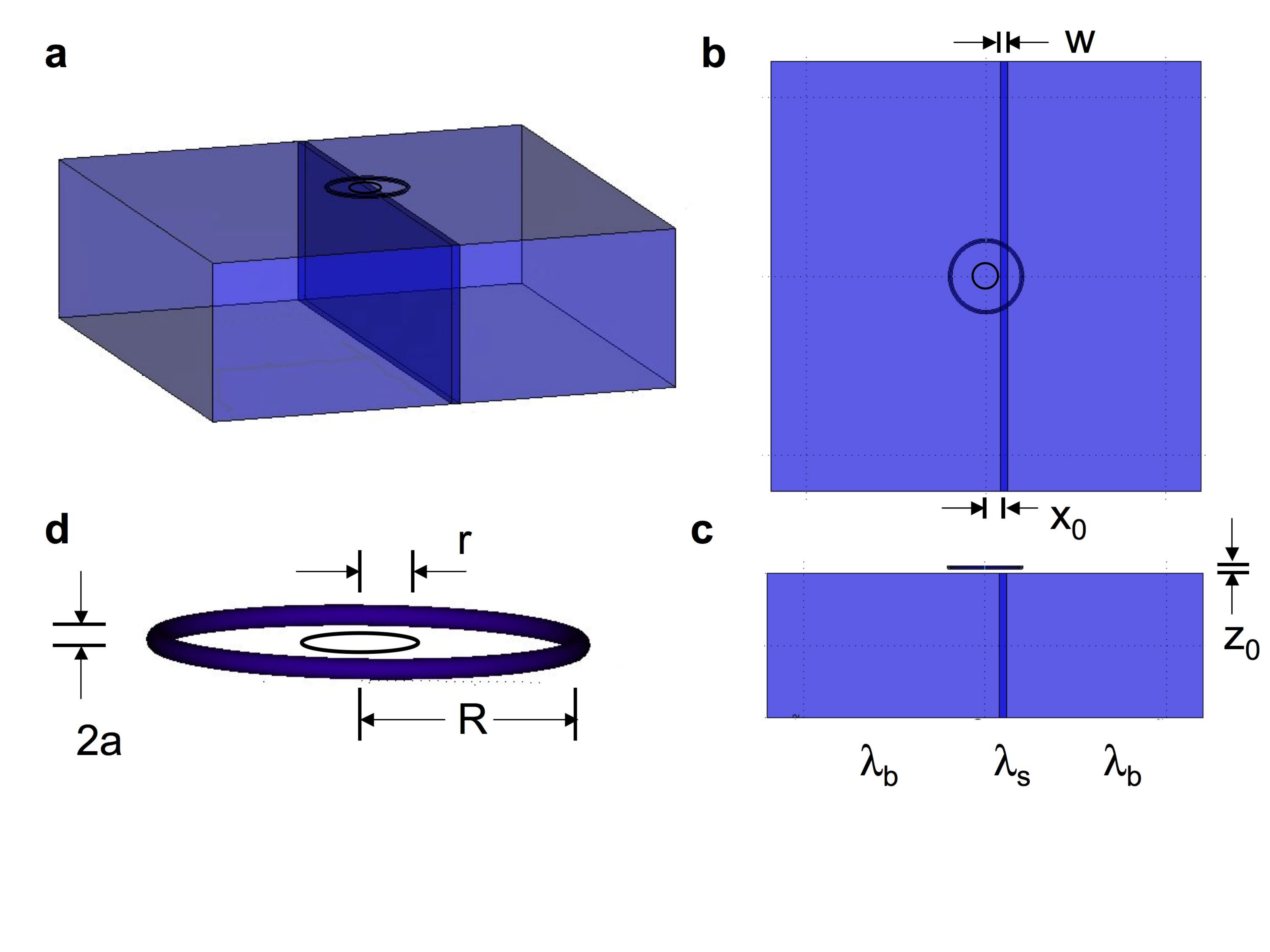}
\caption{Modeling geometry: (a) 3-d rendering; (b) top view; (c) side view; and (d) expanded view of  the field coil. A superconductor occupying the half-space z$<$0 is assumed to have a penetration depth $\lambda_b$ except for a sheet of width $w$ with penetration depth $\lambda_s$ centered at $x=x_0$. The susceptometer field coil, represented by a torus with major radius $R$ and minor radius $a$ oriented with its plane parallel to the surface of the superconductor and centered at $x=0, y=0, z=z_0$, carries a current $I$. The susceptibility $\chi = \Phi / I$ is calculated by integrating the $z$-component of the resultant magnetic field over a circle of radius $r$ centered at ($0,0,z_0$) to obtain the flux $\Phi$. For the modeling in this paper we used $R$=8.85$\mu$m, $z_0/R$=0.17, $a/R$=0.05, $r/R$=0.25. The volume of space modeled was of size 12$\times$12$\times$8 $R^3$.}
\label{fig:comsol_geometry}
\end{figure}

\section{Numerical model}

We believe that the stripes in susceptibility seen in Fig.'s \ref{fig:cross_image} and \ref{fig:cross_ave} are due to reduced penetration depths  associated with the twins. We have not been able to perform an analytical calculation of the change in susceptibility due to a narrow sheet of superconductor with reduced penetration depth. Instead, we numerically solved the coupled Maxwell's and London's equations in the problem with COMSOL, a commercial program that uses finite element methods to solve partial differential equations. The geometry we assumed is shown in Figure \ref{fig:comsol_geometry}. We used COMSOL to numerically solve Laplace's equation $\nabla^2 \vec{H} = 0 $ for $z>0$, and  Londons' equation $\nabla^2\vec{H}=\vec{H}/\lambda^2$ for $z<0$, with $\lambda=\lambda_s$ in the sheet volume $|x-x_0|<w/2$, $z<0$, and $\lambda=\lambda_b$ in the volumes $|x-x_0|>w/2$, $z<0$. The boundary conditions used were continuity of $\vec{H}$ across the plane $z=0$ and the half-planes $x=x_0-w/2$, and $x=x_0+w/2$ ($z<0$), $\vec{H}=0$ on the outside surfaces of the enclosing space, and $H_t = I/2\pi a$, where $H_t$ is the component of $\vec{H}$ tangential to the surface of the torus along the direction of maximum surface curvature. This last boundary condition was derived by using Ampere's law along circles around the torus walls. In general the magnetic fields are not constant along these circles, but this approximation becomes exact as $a \rightarrow 0$. For these simulations we used $a/R=0.05$. Doubling $a$ changed the computed results by only a few percent. The simulations shown here used  meshes generated by an advancing front algorithm with the COMSOL setting $hauto$, which automatically sets several parameters for the meshing,  equal to 4. This resulted in about 600000 degrees of freedom. Using the COMSOL mesh density setting $hauto=4$, with 4 times fewer degrees of freedom, resulted in values for the $z$-component of the magnetic field at ($0,0,z_0$) which were different from those for the more dense mesh by a few percent. For all the modeling presented here we used the denser mesh, which took about 8 minutes per point on a 2.5 GHz Intel Core 2 Duo Mac Book Pro.

\section{Results}

\begin{figure}
\includegraphics[width=6in]{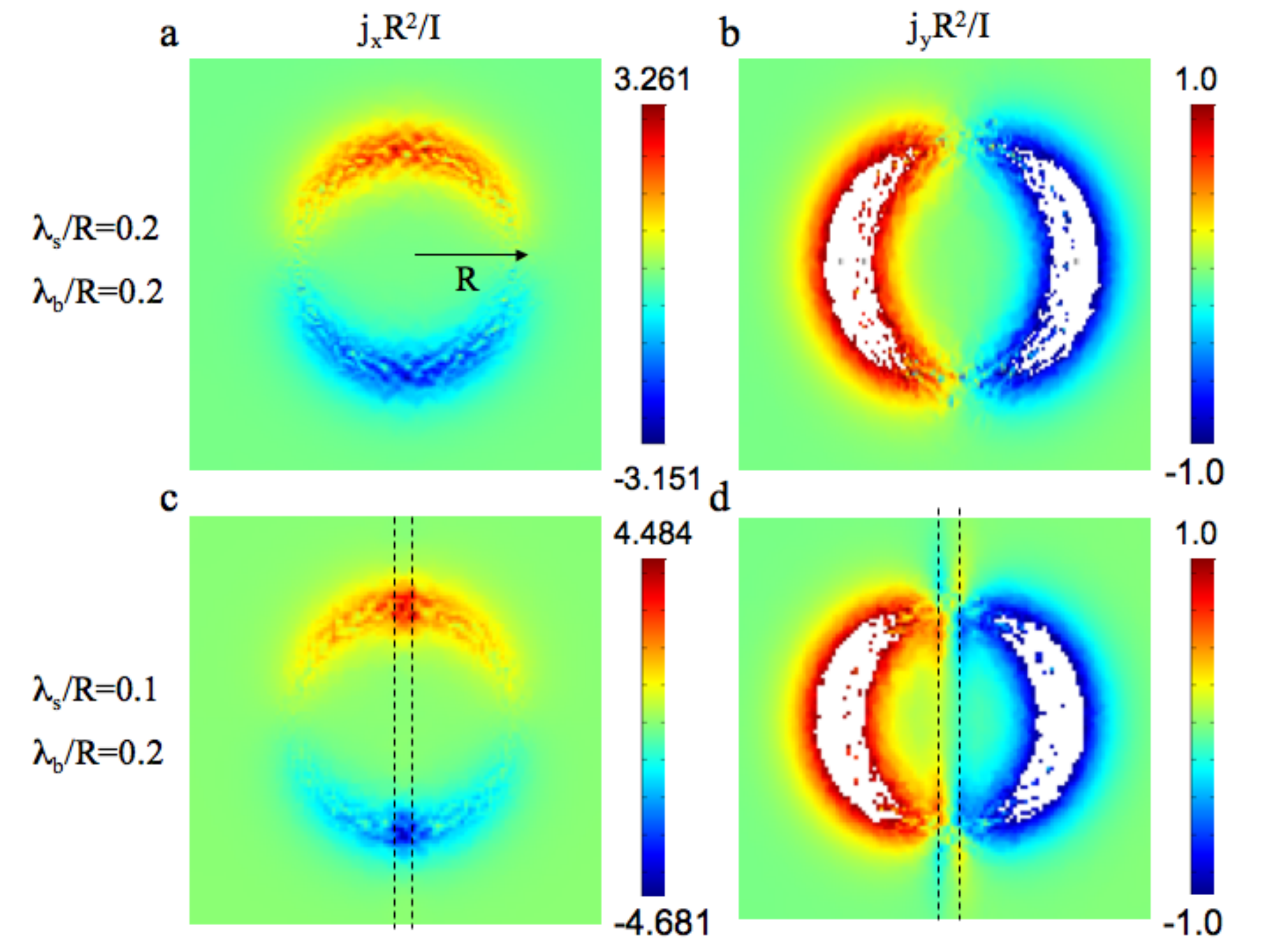}
\caption{Calculated normalized supercurrent densities at $z$=0 for $w/R$=0.2  in the $\hat{x}$ (a,c) and $\hat{y}$ (b,d) directions for the sheet penetration depth $\lambda_s/R$=0.2 and bulk penetration depth $\lambda_b/R$=0.2. (a,b),  and $\lambda_s/R$=0.1, $\lambda_b/R$=0.2 (c,d). The small scale inhomogeneities in the image are due to discretization error. }
\label{fig:js}
\end{figure}

Figure \ref{fig:cross_ave} compares the experimental cross-section from Fig. \ref{fig:cross_image} with our numerical results, obtained by integrating $H_z$, the calculated field in the $\hat{z}$ direction, using $w/R=0.2$, $\lambda_s/R$=0.1, and $\lambda_b/R$=0.12, over the area of the pickup loop for a number of different values of $x_0$.  Noise is visible in the modeling due to the small difference between $\lambda_s$ and $\lambda_b$ required to fit the experimental stripe heights, given the value for $w/R$ chosen. Smaller values of $w/R$ require more mesh elements than can be accommodated by our computer memory. We will discuss how the stripe peak heights scale with the different parameters below. The susceptibility contributions from the 8 stripes visible in the image were added. The fitting parameters in this plot were $\lambda_b$, $\lambda_s$, $w$, the center positions of the stripes and a uniform vertical shift. This figure shows that the stripe peak shapes and widths agree with experiment within the variation from stripe to stripe: The simulated peak heights are $3.43\pm0.01 \Phi_0/A$ while the experimental peak heights are $3.37\pm0.35 \Phi_0/A$. The experimental peak full-widths at half-maximum (FWHM) are $8.92\pm0.62 \mu m$, while the simulated peak widths are $9.34\pm0.82\mu m$. The stripe susceptibility peak widths are limited by experimental resolution rather than their intrinsic widths: Equally good agreement with the experimental peak width can be obtained for any value of $w$ below about 5$\mu$m. Increasing the simulated width to $w/R=0.6$, corresponding to $w=5.3\mu m$, gives a simulated FWHM of $10.07\pm0.75\mu m$, about one standard deviation larger than the experimental peak widths.

Figure \ref{fig:js} shows results for the dimensionless current densities $j_xR^2/I$ and $j_yR^2/I$ (using $\vec{j}=\nabla \times \vec{H}$) in the $\hat{x},\hat{y}$ directions respectively at $z=0$, the surface of the superconductor, induced by the field coil. In the case in which the penetration depth of the sheet is the same as the rest of the superconductor (Fig. \ref{fig:js}a,b, $\lambda_s/R=\lambda_bR$=0.2, ) the field coil induces screening currents which are strongest directly under the ring, and circulate with the opposite sense as the currents in the field coil. If the sheet penetration depth is smaller than that of the bulk (Fig. \ref{fig:js} c,d, $\lambda_s/R$=0.1, $\lambda_b/R$=0.2) there is an additional component of the screening current in the $\hat{x}$ direction concentrated under the field coil, and a more delocalized excess component in the $\hat{y}$ direction. We chose a larger difference between $\lambda_b$ and $\lambda_s$ for this image than for the fit of Fig. \ref{fig:cross_ave} to increase the contrast in the sheet region.

\section{Scaling } 
\begin{figure}
\includegraphics[width=6in]{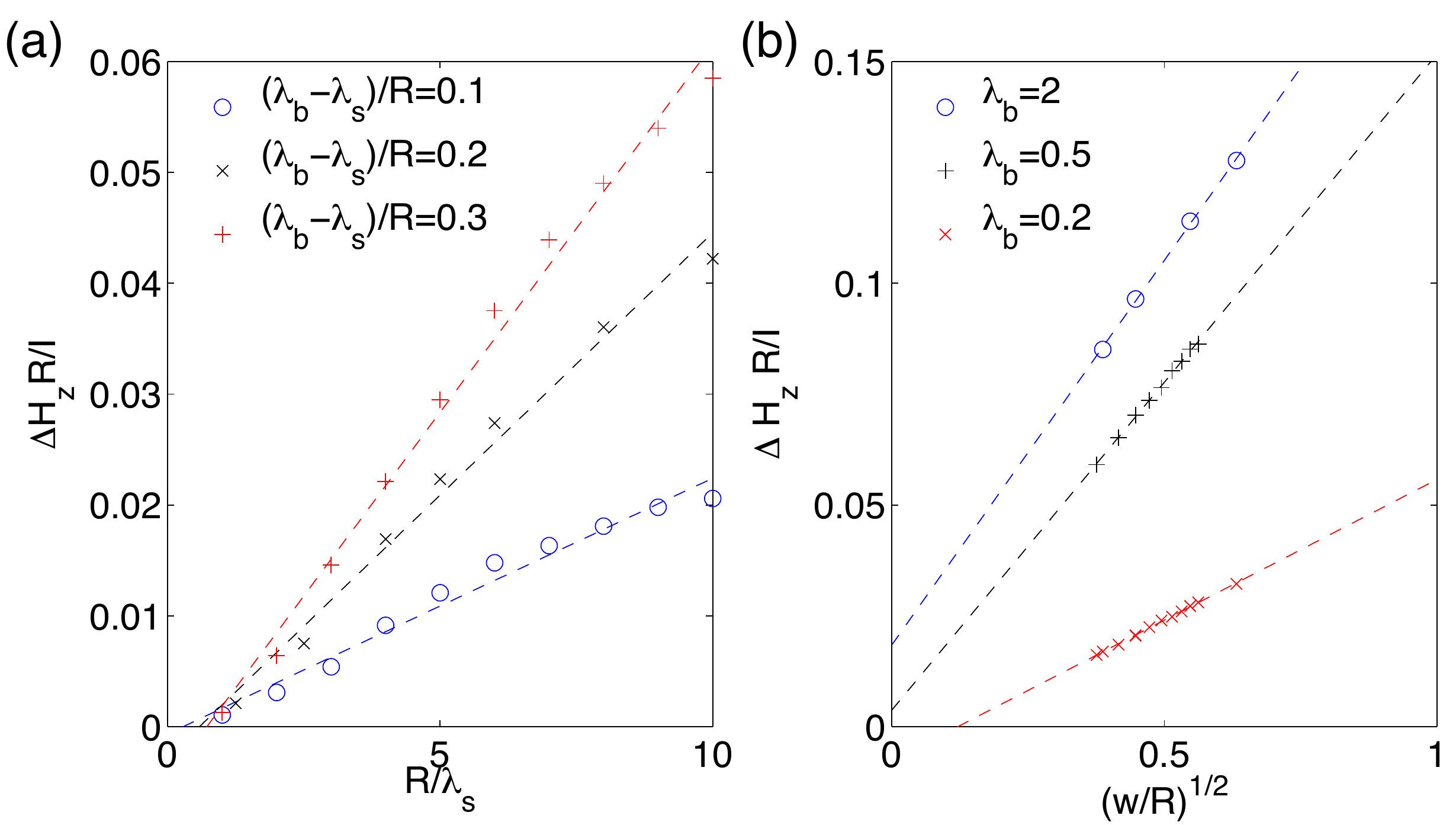}
\caption{Dependence of the stripe peak height $\Delta H_z R/I$ vs. $1/\lambda_s$(a) and $w^{1/2}$ (b), where $\Delta H_z$ is the difference between $x_0/R=0$ and $x_0/R=2$ of the $z$-component of the magnetic field at the center of the field coil induced by the field coil current $I$. The symbols are the numerical simulation and the dashed lines are linear fits.  }
\label{fig:peak_Hz_vs_lambdas_and_w}
\end{figure}

\begin{figure}
\includegraphics[width=4in]{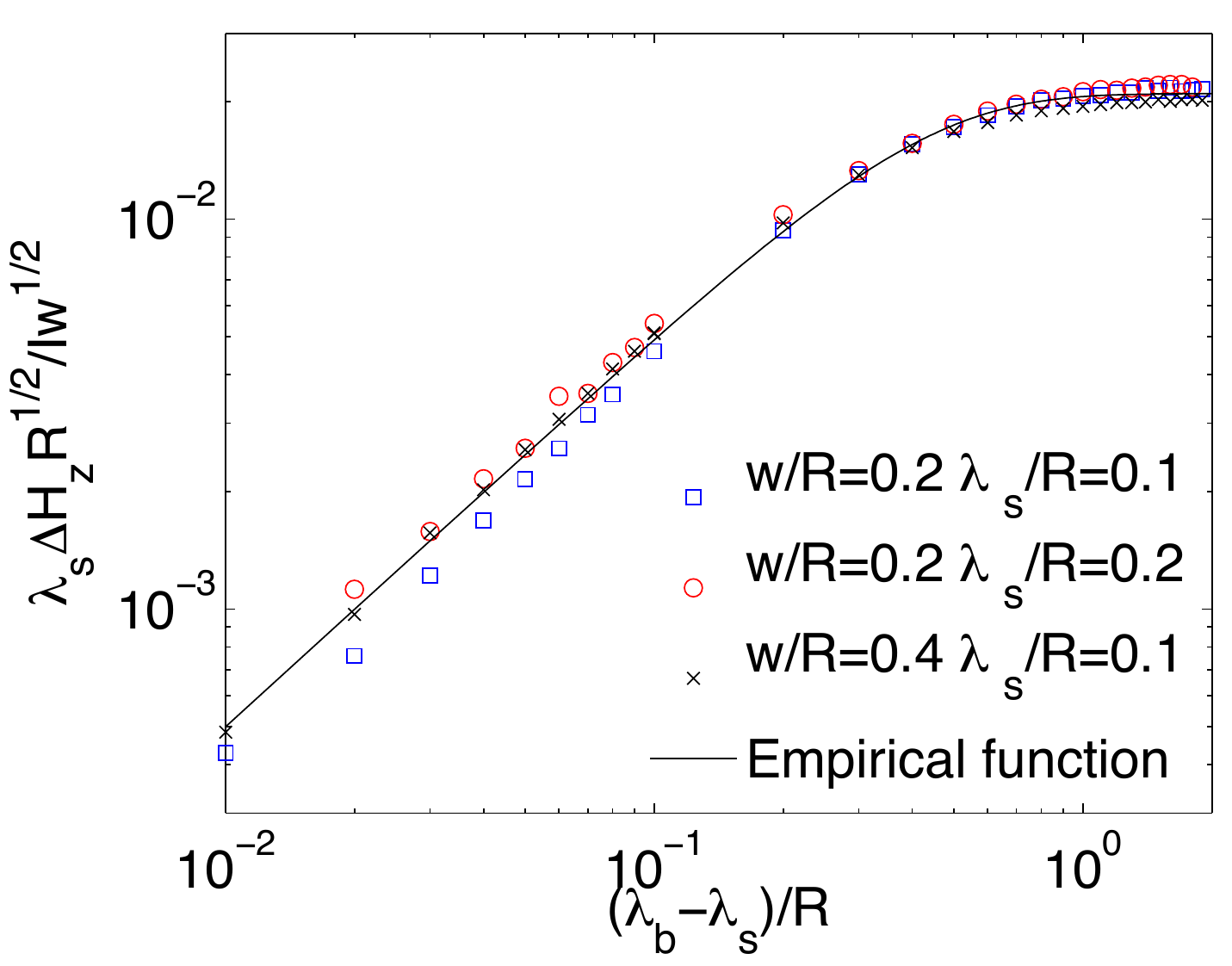}
\caption{Plot of the dimensionless stripe susceptibility peak height $\lambda_s\Delta H_zR^{1/2}/Iw^{1/2}$ vs $(\lambda_b-\lambda_s)/R$ for various values of the reduced sheet width $w/R$ and sheet penetration depth $\lambda_s/R$. For large values of $\lambda_b-\lambda_s$ the susceptibility peak height is nearly independent of $\lambda_b$, indicating that the susceptibility peak height is proportional to $w^{1/2}/\lambda_s$, and therefore proportional to the square root of the two-dimensional sheet Cooper pair density $N_s$. The susceptibility peak height approaches zero linearly as $\lambda_b \rightarrow \lambda_s$. The solid line is the empirical function Eq. \ref{eq:scaling}, used for the modeling of the temperature dependence of the stripe amplitude below.}
\label{fig:scaling}
\end{figure}

As discussed above, the experimental width of the stripes is resolution limited.  The sheets of enhanced superfluid density could be as narrow as the coherence length ($\sim$3-4 nm \cite{yin2009}). We therefore performed simulations  to see how the predicted results scaled as $w$ became small. The results are shown in Figures \ref{fig:peak_Hz_vs_lambdas_and_w}-\ref{fig:scaling}. In Figure \ref{fig:peak_Hz_vs_lambdas_and_w} we plot the dimensionless susceptibility peak height $\Delta H_z R/I$, the difference between the $z$-component of the magnetic field at the center of the field coil induced by the current $I$ for $x_0=0$ minus that at $x_0=2R$, as a function of $R/\lambda_s$ (a) and vs. $(w/R)^{1/2}$ (b). The dashed lines in this figure are linear fits. The  susceptibility peak heights are roughly proportional to $1/\lambda_s$, with a slope which is proportional to $\lambda_b-\lambda_s$ (a), and roughly proportional to $w^{1/2}$ (b). The two figures Fig. \ref{fig:peak_Hz_vs_lambdas_and_w}a and \ref{fig:peak_Hz_vs_lambdas_and_w}b indicate that the susceptibility stripe amplitudes scale as $w^{1/2}/\lambda_s$, proportional to the square root of the two-dimension sheet Cooper pair density. In Figure \ref{fig:scaling} we have plotted the dimensionless normalized susceptibility peak height $\lambda_s H_z R^{1/2}/Iw^{1/2}$ versus $(\lambda_b-\lambda_s)/R$, the reduced difference in penetration depths between the stripe and the bulk. The scaling works reasonably well over the range of parameters chosen. The solid line in Figure \ref{fig:scaling} is the empirical relation
\begin{equation}
\frac{\lambda_s \Delta H_z R^{1/2}}{I w^{1/2}} = \alpha \tanh \left ( \beta \frac{\lambda_b - \lambda_s}{R} \right )
\label{eq:scaling}
\end{equation}
with $\alpha=0.021$ and $\beta=2.3796$. We use this relation in the modeling that follows. 

\section{Temperature dependence}

\begin{figure}
\includegraphics[width=4in]{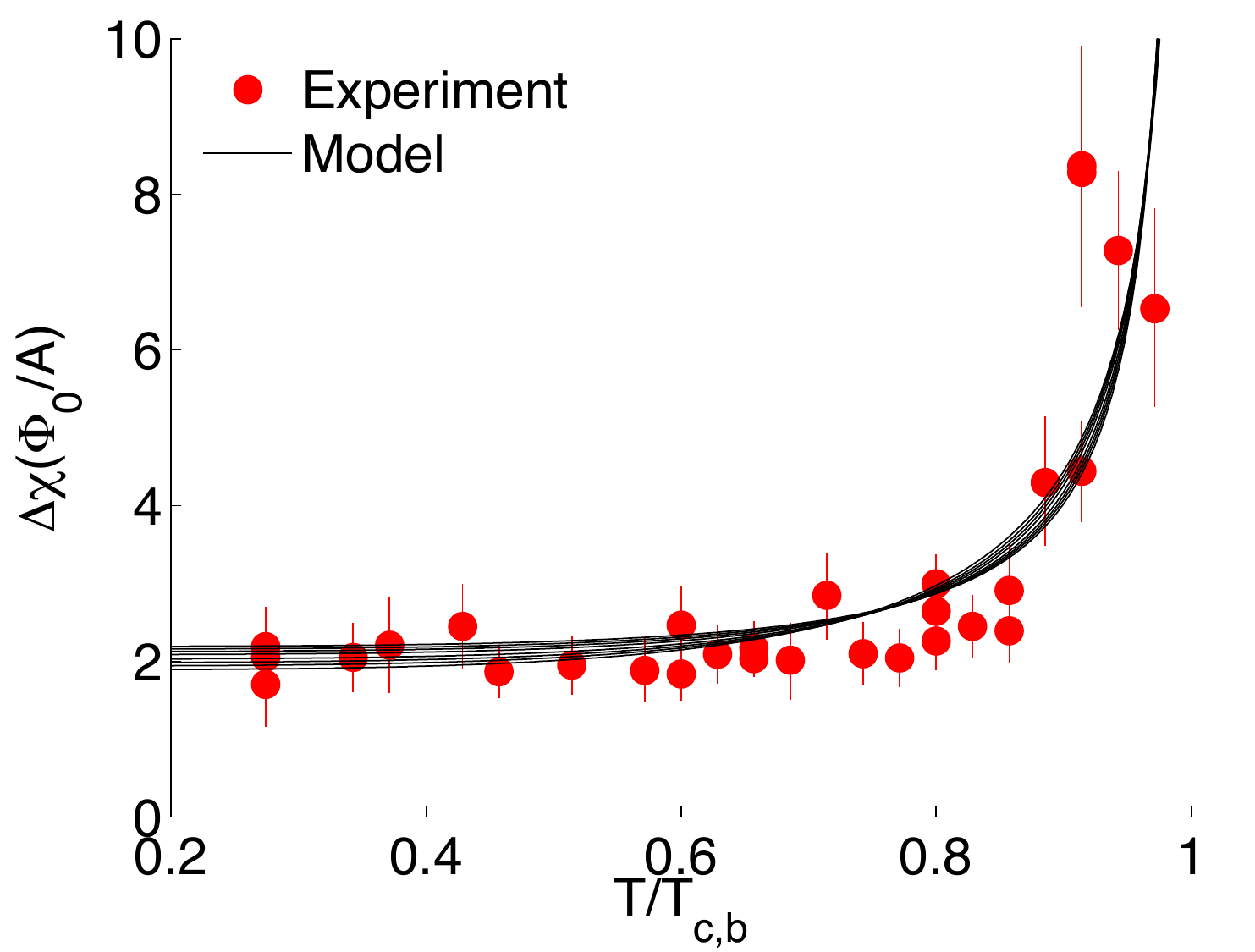}
\caption{Plot of the temperature dependence of the stripe amplitude $\Delta \chi(T)$ (symbols). The solid lines are best fits to the data using the empirical formula Eq. \ref{eq:scaling}  and the two-fluid temperature dependence (Eq. \ref{eq:p}, $p$=4) for the penetration depths, and different critical temperatures $T_{c,s}$ and $T_{c,b}$ for the sheet and bulk respectively, for widths $w/R=0.5, 0.2, 0.1, 0.05, 0.02, 0.01, 0.005,0.002$, corresponding to the sheet width  $w$ between 4.4$\mu$m and 17.6nm. The best fit values and uncertainties for $\lambda_s$ and $T_{c,s}/T_{c,b}$ are plotted in Figure \ref{fig:lambdas_ts_bounds}. }
\label{fig:t_dependence}
\end{figure}

Kalisky {\it et al.} found that as the sample temperature approached $T_c$ the stripe peak amplitudes became larger. Experimental data on an underdoped (x=5.1\%) sample is shown as the symbols in Figure \ref{fig:t_dependence}. The solid symbols in this figure were obtained by fitting an image with multiple stripes to our numerical model, using the procedure of Fig. \ref{fig:cross_ave}, but with a single vertical scaling factor, for a number of sample temperatures. The vertical error bars were obtained by varying the scaling factor from its optimal value until the $\chi^2$ between fit and experiment doubled. We estimate that the errors in temperature are smaller than the symbol widths. Our modeling shows that if the sheet and bulk penetration depths had the same temperature dependence there would be no change in the stripe amplitude until the bulk penetration depth became comparable to the radius of the field coil - very close to T$_c$. We have considered two scenarios for a difference in temperature dependences in the stripe amplitudes: 1) The bulk and sheet critical temperatures are the same, but the penetration depths have different temperature dependences below T$_c$. Such a difference could occur, for example, if the bulk is a weak coupling superconductor, and the sheet is a strong coupling one, or visa versa.  In addition, such a difference could occur if the pairing symmetry is different in the sheet than in the bulk. The temperature dependence of the penetration depth can be parameterized as \cite{prozorov2006}
\begin{equation}
\lambda(T) = \lambda(0)/\sqrt{1-(T/T_c)^p}
\label{eq:p}
\end{equation}
where, for example, good fits to the BCS predictions can be obtained using $p$=2 for $s$-wave and $p$=4/3 for $d$-wave pairing symmetries \cite{prozorov2006}. We fit the experimental data of Fig. \ref{fig:t_dependence} using the Eq. \ref{eq:scaling} and Eq. \ref{eq:p} using $\lambda_s$, $p_s$, and $p_b$ (the $p$ exponents for the sheet and bulk respectively) as the three fitting parameters. In all cases the fits gave unreasonably large values for the $p$'s. For example, the best fit assuming $w/R$=0.2, $\lambda_b$=0.04 gave $p_s$=34.2 and $p_b$=17.9. It therefore seems unlikely that the first scenario, in which the twin boundaries and bulk have the same critical temperature, can explain the data.

\begin{figure}
\includegraphics[width=4in]{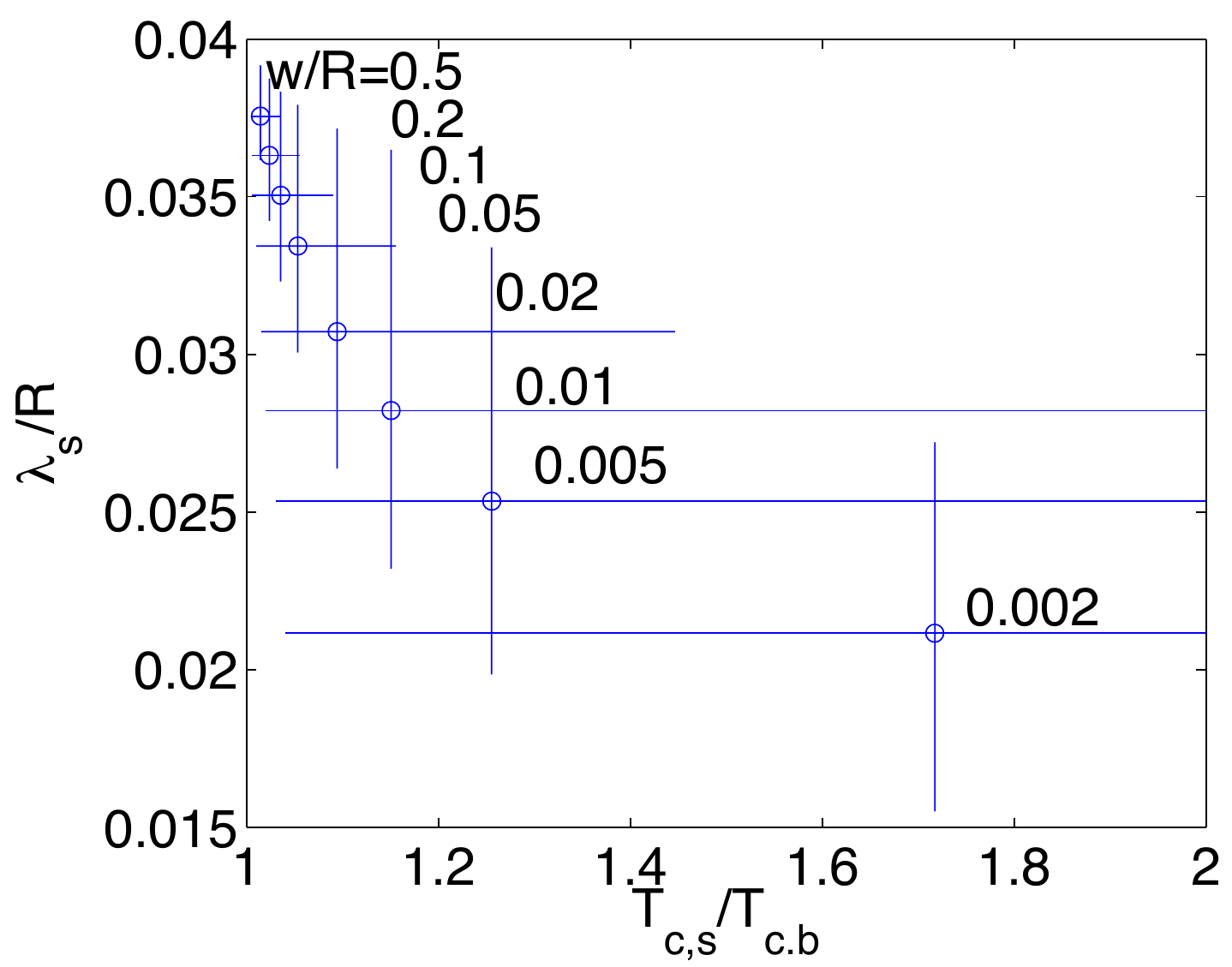}
\caption{Best fit values and uncertainties (using a doubling of the best fit value for $\chi^2$ as a criterion) for the parameters $\lambda_s/R$ and $T_{c,s}/T_{c,b}$ obtained by fitting the data of Figure \ref{fig:t_dependence}, for various assumed values for the sheet width $w$, assuming $\lambda_b(T=0)/R=0.04$.}
\label{fig:lambdas_ts_bounds}
\end{figure}

2) On the other hand, good fits to the data can be obtained by assuming different critical temperatures for the bulk and sheet. The solid lines in Figure \ref{fig:t_dependence} are a superposition of fits of Eq. \ref{eq:scaling} to the data, assuming that both the sheet and bulk penetration depths follow the two-fluid temperature dependence ($p=4$) in Eq. \ref{eq:p},
assuming two different $T_c$'s for the bulk and the sheet, for various values of $w/R$, and $\lambda_b(T=0)/R$=0.04 (corresponding to $\lambda_b=0.35\mu m$\cite{luan2009}). The quality of the fits is not significantly different for different values of the sheet width $w$ within the physically reasonable range. Similar results, with slightly different values for $T_{c,s}/T_{c,b}$, are obtained using $p=2$ (BCS) or $p=4/3$ ($d$-wave) in Eq. \ref{eq:p} for the fits. Figure \ref{fig:lambdas_ts_bounds} displays the best fit values and error bars (using a doubling of the best-fit $\chi^2$ as the criterion) for $\lambda_s/R$ and $T_{c,s}/T_{c,b}$ from these fits. In this modeling we assume that $w$ is independent of temperature. Since the stripe amplitudes scale like $w^{1/2}$, any $w$ temperature dependence can be neglected relative to the temperature dependence of $\lambda_b-\lambda_s$.

\begin{figure}
\includegraphics[width=4in]{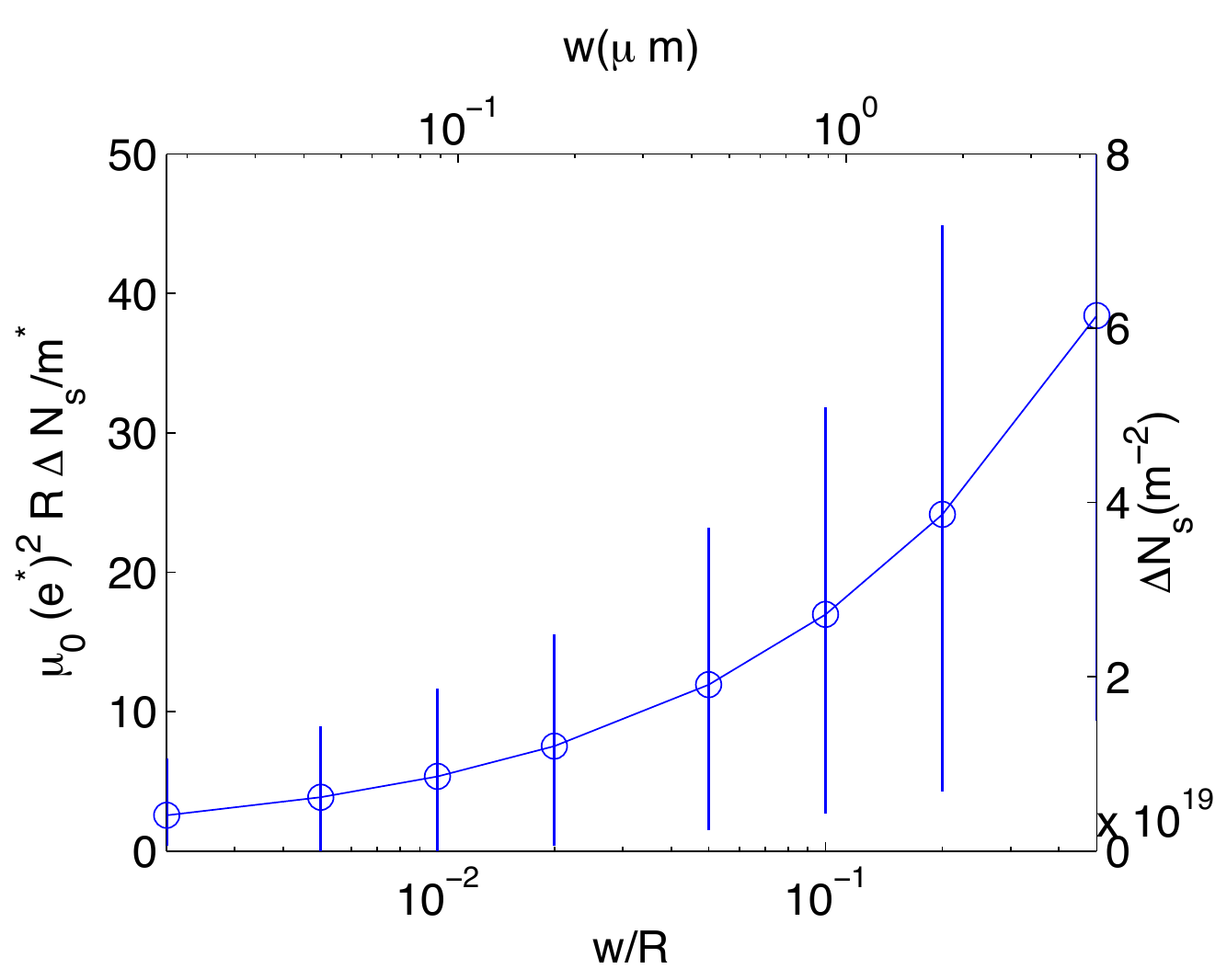}
\caption{Best-fit values, and upper and lower bounds for the enhancement of the two-dimensional sheet Cooper pair density as a function of sheet width from fits to experimental susceptibility stripe amplitudes. }
\label{fig:Ns_bounds}
\end{figure}

\section{Discussion}

Our analysis above indicates that the twin boundaries in underdoped Ba(Fe$_{1-x}$Co$_x$)$_x$As$_2$ have a shorter penetration depth than the bulk. A shorter penetration depth is usually associated with a higher superfluid density, although the relation between the two quantities could be modified by fluctuations \cite{roddick1995}. We assume for simplicity the London relation to obtain an estimate for the excess 2-dimensional sheet Cooper pair density $\Delta N_s$ 
\begin{equation} 
\Delta N_s=w\frac{m^*}{\mu_0(e^*)^2}\left (\frac{1}{\lambda_s^2}-\frac{1}{\lambda_b^2} \right )
\label{eq:london}
\end{equation}
where $m^*$ is the Cooper pair mass (assumed to be twice the bare mass of the electron) and $e^*$=2e is the Cooper pair charge. From Fig. \ref{fig:Ns_bounds} we estimate that the excess 2-dimensional sheet Cooper pair density is $10^{19}  m^{-2}  < \Delta N_s < 10^{20} m^{-2}$. 

Our analysis  also indicates that the sheets have a higher critical temperature than the bulk, although stripes have not yet been observed above the bulk critical temperature, possibly because of superconducting fluctuations or local variations in the bulk T$_c$. 
%One explanation for this could be that at temperatures above the bulk critical temperature but below the sheet critical temperature the superconductivity on the twin boundaries is 2-dimensional, in which case phase fluctuations would destroy diamagnetic shielding at temperatures above the Kosterlitz-Thouless \cite{kosterlitz1973} temperature
%\begin{equation}
%T_{KT} \approx \frac{\Phi_0^2\mu_0 e^2N_s}{32 k_B\pi^2 m},
%\label{eq:kt}
%\end{equation}
%where $\Phi_0=h/2e$ is the superconducting flux quantum and $k_B$ is Boltzman's constant. Using the estimates for the sheet Cooper pair density above, we find $4\times 10^{-4} K < T_{KT} < 4 \times 10^{-3}$ K: the Kosterlitz-Thouless temperature is well below T$_c$. This could explain why enhanced susceptibility on the twins was not observed above the bulk critical temperature.

Kalisky {\it et al.} also found that vortices were not pinned on the sheets, and that it was not possible to drag vortices across the sheets. The energy required to form a vortex in a superconductor is given approximately by\cite{tinkham}:
\begin{equation}
E_v \approx \frac{\Phi_0^2 L}{4\pi\mu_0\lambda^2}\ln \kappa,
\label{eq: vortex_energy}
\end{equation}
where $\kappa \approx \lambda/\xi \approx 140$ \cite{ishida2009}, and $L$ is the vortex length. From the estimates of Figure \ref{fig:lambdas_ts_bounds} we find $1\times10^{-17}J < \Delta E_v < 4\times10^{-16}J$ for the difference in energy of the vortex on vs. off the sheet,  assuming the crystal thickness $L=10\mu m$. These energies are much larger than $k_B T_c$. These are very rough estimates, because it may well be that the sheet width is much smaller than the penetration depth, in which case much of the vortex field energy is outside of the sheet. We estimate from dragging experiments with an MFM tip that a force of 40 pN at 5K and 6pN at 14K was not enough to make a vortex cross a sheet. Using $w$ as a characteristic length, the maximum and minimum excess vortex energies on the sheet correspond to forces $F_s \approx \Delta E_v/w$ of $1.4\times 10^{-9}N > F_s >  9.2 \times 10^{-11} N$, easily large enough to explain the inability to drag vortices across the twin boundaries.

Subsequent to the work of Kalisky {\it et al.} \cite{kalisky2009}, Prozorov {\it et al.} \cite{prozorov2009}  noted an enhancement of the critical current of slightly underdoped single crystals of Ba(Fe$_{1-x}$Co$_x$)$_2$As$_2$, which they associated with twin boundaries. The measurements of Kalisky {\it et al.} provide two mechanisms for this critical current enhancement: 1) Enhanced superfluid density along the twin boundaries provides channels with enhanced depairing current densities, and 2) The enhanced superfluid densities in the twin boundaries provides barriers to transverse vortex motion. Our modeling shows that the latter effect can be substantial.

\section{Conclusions}

In conclusion, we have used finite element methods to numerically solve Maxwell's and London's equations for the problem of enhanced susceptibility from a narrow sheet of superconductor with reduced penetration depth embedded in a bulk superconductor in a geometry appropriate for scanning SQUID susceptometry measurements. We find good agreement between our modeling and experiment for the lineshape for cross-sections across stripes in enhanced susceptibility measurements on underdoped samples of the pnictide superconductor Ba-122. By scaling our simulations and comparing the results with experiment we obtain estimates of the enhanced Cooper pair sheet density on the sheet. The barrier energies and forces we estimate are large enough to explain the observed lack of pinning on the sheets, and the experimental inability to drag vortices across the sheets.

\section*{Acknowledgements}
We would like to thank T. Lippman for assistance in acquiring and installing COMSOL, and S. Kivelson and V. Kogan for useful conversations. This work was supported by the Department of Energy DOE Contract No. DE-AC02-76SF00515, by NSF Grant No. PHY-0425897, and by the US-Israel Binational Science Foundation.  BK acknowledges the support of the L'Oreal USA Fellowships For Women in Science program.

\end{document}